\documentclass[aps,pra,twocolumn,nofootinbib]{revtex4-2}
\usepackage{amsmath,amssymb,amsthm}
\usepackage{bm,bbm}

\usepackage{graphicx,subfigure}

\usepackage[bookmarks=false]{hyperref}
\hypersetup{colorlinks=true,citecolor=blue,linkcolor=red,%
urlcolor=blue,pdfstartview=FitH,bookmarksopen=true}

\usepackage[T1]{fontenc}
\usepackage[osf,sc]{mathpazo}
\usepackage{dsfont}

\newcommand{\ket}[1]{\mbox{$ | #1 \rangle $}}
\newcommand{\bra}[1]{\mbox{$ \langle #1 | $}}

\newcommand{\tr}{\mathrm{tr}}

\newcommand{\cC}{\mathcal{C}}

\newtheoremstyle{note}
  {\topsep/2}              	
  {\topsep/2}            	
  {}                        
  {\parindent}             	
  {\itshape}                
  {.---}                    
  {0pt}                     
  {\thmname{#1}\thmnumber{ \itshape#2}\thmnote{ (#3)}} 

\newtheorem{theorem}{Theorem}

\newtheorem{proposition}[theorem]{Proposition}

\theoremstyle{definition}

\theoremstyle{remark}

\begin{document}
\title{Operational detection of entanglement via quantum designs}

\author{Xin Yan}
\affiliation{Key Laboratory of Advanced Optoelectronic Quantum Architecture and Measurement of Ministry of Education, School of Physics, Beijing Institute of Technology, Beijing 100081, China}

\author{Ye-Chao Liu}
\email{yechaoliu1994@outlook.com}
\affiliation{Key Laboratory of Advanced Optoelectronic Quantum Architecture and Measurement of Ministry of Education, School of Physics, Beijing Institute of Technology, Beijing 100081, China}

\author{Jiangwei Shang}
\email{jiangwei.shang@bit.edu.cn}
\affiliation{Key Laboratory of Advanced Optoelectronic Quantum Architecture and Measurement of Ministry of Education, School of Physics, Beijing Institute of Technology, Beijing 100081, China}

\date{\today}
%

\begin{abstract}
From an operational point of view, we propose several new entanglement detection criteria using quantum designs.
These criteria are constructed by considering the correlations defined with quantum designs.
Counter-intuitively, the criteria with more settings are exactly equivalent to the corresponding ones with the minimal number of settings, namely the symmetric informationally complete positive operator-valued measures (SIC POVMs).
Fundamentally, this observation highlights the potentially unique role played by SIC POVMs in quantum information processing.
Experimentally, this provides the minimal number of settings that one should choose for detecting entanglement.
Furthermore, we find that nonlinear criteria are not always better than linear ones for the task of entanglement detection.
\end{abstract}

\maketitle
%

\section{Introduction}\label{sec:Intro}
Entanglement \cite{Guehne.Toth2009,Horodecki.etal2009} is one of the most notable characteristics of quantum theory as compared to classical theory.
Apart from its fundamental importance, quantum entanglement also plays a vital role in various tasks in quantum information processing.
Then, a basic yet crucial question to ask is how to determine whether a given quantum state is entangled or not.
Although considerable results have been obtained, a universal method for checking entanglement is still not available.

A bipartite quantum state $\rho_{AB}$ is called separable if it can be written as a convex combination of product states, i.e.,
\begin{equation}
  \rho_{AB}=\sum_{k}p_{k}\ket{a_{k}}\bra{a_{k}}\otimes\ket{b_{k}}\bra{b_{k}}\,,
\end{equation}
where the coefficients $\{p_{k}\}$ form a probability distribution with ${p_{k}\geqslant 0}$ and ${\sum_{k}p_{k}=1}$; Otherwise, $\rho_{AB}$ is entangled.
The well-known positive partial transpose (PPT)  criterion \cite{Peres1996,Horodecki.eta1996} is both necessary and sufficient for detecting entanglement for the simple ${2\times2}$ and ${2\times3}$ systems, but not for higher dimensions.
For instance, there exist the so-called bound entangled states \cite{Horodecki1998} which are PPT and nondistillable.
An equivalent generalization of PPT, namely the entanglement witness \cite{Guehne.Toth2009,Horodecki.etal2009,Horodecki.eta1996}, is a universal method for detecting an arbitrary entangled state $\rho$ with ${\tr(W\rho) < 0}$ where $W$ is a suitably-chosen (but may not be unique) Hermitian observable.

While the construction of an entanglement witness can be difficult sometimes, a number of non-universal but easily applicable methods have been proposed.
The most prominent example is the computable cross-norm or realignment (CCNR) criterion \cite{Horodecki.eta2006,Rudolph2005,Albeverio2003}, which is based on the correlations of local orthogonal observables (LOOs).
Then, the local uncertainty relation (LUR) \cite{Hofmann2003,Otfried2006} criterion extends the CCNR by adding extra nonlinear terms which can also be regarded as a natural nonlinear entanglement witness for CCNR.
Another class of linear correlation-based entanglement criteria is constructed using the normalized symmetric informationally complete positive operator-valued measures (SIC POVMs) \cite{Renes2004}, dubbed as the ESIC criterion \cite{Shang.etal2018}; see also Refs.~\cite{chen2015,li2020entanglement}.
Very recently, a general approach for checking separability is taken by considering the linear correlations of specific operators in Refs~\cite{SarbickiGniewomir2020,Sarbicki2020}, which incidentally recovers both the CCNR and the ESIC criteria.

Apart from SIC POVMs, another extension for entanglement detection is via mutually unbiased bases (MUBs) \cite{William1989,chen2013,tao2015,Spengler2012,Bennett1999,Chen2014}, both of which are special cases of quantum $2$-designs.
For instance, a measurement-device-independent entanglement detection method is discussed in Ref.~\cite{Bae2019} using SIC POVMs and MUBs.
Recently, depending on the random moments calculated from quantum designs, method for characterizing multipartite entanglement is proposed \cite{Ketterer2019,Ketterer2020}.
Generally speaking, however, the random moments are not directly measurable in experiments.
Therefore, here we focus on the correlations defined via quantum designs from an operational point of view.
As quantum designs correspond to a single generalized measurement, the correlations can be directly obtained in one go in experiments, rather than one by one as in other criteria like the CCNR. 
Specifically, we propose several new entanglement criteria using quantum $2$-designs by extending the CCNR, ESIC, and LUR criteria.

This paper is organized as follows.
We first briefly review the concept of quantum designs in Sec.~\ref{sec:Design}, with a special emphasis on the discussion of SIC POVMs.
In Sec.~\ref{sec:EntCriteria}, upon recalling some well-known entanglement criteria including the CCNR, ESIC, and LUR criteria, we construct several new criteria using quantum designs.
Then these criteria are tested using various bipartite entangled states in Sec.~\ref{sec:Appl},
and we conclude in Sec.~\ref{sec:Summary}.

\section{Quantum Designs}\label{sec:Design}
Design is an important mathematical concept which can be used to imitate uniform averages over certain groups, which in turn can be regarded as a pseudorandom process.
Designs are denominated either unitary or spherical, hinging on which group one chooses.
For qubit systems, the local measurement settings can be characterized over the Bloch sphere, then it is more convenient to use spherical designs than unitary designs.
A spherical $t$-design is a collection of points on the unit sphere for which the $t$th-order polynomials can be averaged over to obtain the same value as that integrating over the surface with certain measures.
Formally, a probability distribution over the set of quantum states $(p_i,\ket{\phi_i})$ is a quantum spherical $t$-design if
\begin{eqnarray}
  \sum_i p_i (\ket{\phi_i}\bra{\phi_i})^{\otimes t}=\int_{\psi} (\ket{\psi}\bra{\psi})^{\otimes t}\mathrm{d}\psi\,,
\end{eqnarray}
where the integral over \ket{\psi} is taken over the Haar measure on the unit sphere \cite{Ambainis.etal2007}.

In this work, we choose ${t=2}$ in particular to focus on the investigation of bipartite entanglement.
We can associate the complete set of points of a spherical $2$-design as a measurement $\{\widetilde{\Pi}_k\}_{k=1}^N$, where $N$ denotes the number of settings. 
Then, a quantum state $\rho$
can be reconstructed as \cite{Slomczynski2020}
\begin{equation}\label{rho_2de}
\rho=(d+1)\frac{N}{d}\sum_{k=1}^N \widetilde{p}_k\widetilde{\Pi}_k-\openone\,,
\end{equation}
where $d$ is the dimension, and the probability of obtaining the outcome $\widetilde{\Pi}_k$ is given by the Born rule
\begin{equation}
\widetilde{p}_k=\tr\bigl(\rho\widetilde{\Pi}_k\bigr)=\bigl\langle\widetilde{\Pi}_k\bigr\rangle\,.
\end{equation}
The following inequality can be derived directly from Eq.~\eqref{rho_2de}, such that
\begin{equation}\label{widetilde_p^2}
\sum_{k=1}^N \widetilde{p}_k^2=\frac{d\bigl[1+\tr(\rho^2)\bigr]}{(d+1)N}\leqslant\frac{2d}{N(d+1)}\,,
\end{equation}
where the upper bound is saturated when $\rho$ is pure.

For simplicity and later use, let’s rewrite
\begin{equation}
  \Pi_k=\sqrt{\frac{N(d+1)}{2d}}\widetilde{\Pi}_k
\end{equation}
as the normalized version of the quantum $2$-design.
Then, the constraint in Eq.~\eqref{widetilde_p^2} can be reformulated as 
\begin{equation}\label{eq:probS}
  \sum_{k=1}^N p_k^2 \leq 1\,,
\end{equation}
where ${p_k=\tr(\rho\Pi_k)=\langle\Pi_k\rangle}$ for a given state $\rho$, and the equality is achieved when $\rho$ is pure.

Among the typical examples of quantum $2$-designs are SIC POVMs and MUBs \cite{William1989,chen2013,tao2015}.
Both of them have been demonstrated being useful for entanglement detection \cite{Shang.etal2018,Spengler2012,Bennett1999,Chen2014}.
Here we look at SIC POVMs in specific.
A SIC POVM in dimension $d$ comprises of $d^2$ subnormalized projectors $\ket{\psi_k}\bra{\psi_k}/d$ with equal pairwise fidelity, such that
\begin{equation}
|\langle\psi_i|\psi_j\rangle|^2=\frac{d\delta_{ij}+1}{d+1}\,,\quad i,j=1,2,...,d^2\,.
\end{equation}
One can check that the normalized version of the SIC POVM that satisfies the condition in Eq.~\eqref{eq:probS} takes on the form
\begin{equation}\label{eq:SIC}
   E_k=\sqrt{\frac{d+1}{2d}}\ket{\psi_k}\bra{\psi_k}\,.
\end{equation}
Although being widely believed and numerically supported, the existence of SIC POVMs in any finite dimension remains as an open problem \cite{ZaunerThesis2011}.
For a recent review, see Refs.~\cite{Fuchs.etal2017, Ling.etal2006, Medendorp.etal2011, Bent.etal2015}.

\section{Correlation-based entanglement criteria}\label{sec:EntCriteria}
As discussed early, entanglement criteria constructed using correlations are particularly relevant for easier experimental realizations.
To be specific, in this work we first re-investigate the CCNR and ESIC criteria \cite{Shang.etal2018} which are linear, as well as the LUR criterion \cite{Otfried2006} which is nonlinear.
Then, these criteria are extended straightforwardly by utilizing quantum $2$-designs, and the question whether the new criteria thus obtained are improved or not naturally follows.

\subsection{Linear criteria}
Consider a bipartite quantum state $\rho_{AB}$ with the dimension ${d=d_A\times d_B}$, where $d_A$ and $d_B$ represent the local dimensions of the subsystems ${\rho_A=\tr_B(\rho_{AB})}$ and ${\rho_B=\tr_A(\rho_{AB})}$ respectively.
Let $\{M_k^A\}_{k=1}^{K_A}$ and $\{M_k^B\}_{k=1}^{K_B}$ denote the local operations acting on the two subsystems.
Then, the linear correlation matrix between these two measurements can be written as
\begin{equation}
  [\cC]_{ij}=\langle M_{i}^A\otimes M_{j}^B\rangle=\tr\bigl(\rho_{AB}M_{i}^A\otimes M_{j}^B\bigr),
\end{equation}
the size of which is $K_A\times K_B$.
Then, we have the following proposition.
\begin{proposition}\label{pro1}
Let $\{M^A\}$ and $\{M^B\}$ be the properly normalized local measurements acting on the bipartite state $\rho_{AB}$.
If $\rho_{AB}$ is separable, then
\begin{equation}
  ||\cC||_{\tr}\leqslant 1
\end{equation}
has to hold; Otherwise, it is entangled.
The symbol $||\!\cdot\!||_{\tr}$ denotes the trace norm.
\end{proposition}

The proof can be found in Refs.~\cite{Shang.etal2018,SarbickiGniewomir2020}.
Depending on the local measurements one chooses, different entanglement criteria can be derived from Proposition~\ref{pro1}. 
For instance, if $\{M^A\}$ and $\{M^B\}$ are LOOs,
we get the CCNR criterion \cite{Horodecki.eta2006,Rudolph2005,Albeverio2003}.
The LOOs can be found by invoking the Schmidt decomposition (assuming $d_A\leqslant d_B$), such that
\begin{equation}
 \rho_{AB}=\sum_{k=1}^{d_A^2} \lambda_k G_{k}^{A}\otimes G_{k}^{B},
\end{equation}
where ${\lambda_k=\langle G_k^A\otimes G_k^B\rangle}$ are the Schmidt coefficients.
It is easy to check that the set of orthonormal bases of the Hermitian observables $\{G_k^A\}$ and $\{G_k^B\}$ fulfill the conditions
\begin{equation}
   \tr(G_k^A G_l^A)=\tr(G_k^B G_l^B)=\delta_{kl}\,,
\end{equation}
and
\begin{equation}
  \sum_k(G_k^A)^2=d_A\openone\,,\quad\sum_k(G_k^B)^2=d_B\openone\,.
\end{equation}
Hence, an equivalent form of the CCNR criterion is given by
\begin{equation}
    \sum_k\lambda_k\leqslant 1\,,
\end{equation}
as the Schmidt coefficients $\lambda_k$s happen to be the singular values of $\cC$.

If, instead, one chooses the local measurements to be the normalized SIC POVMs as in Eq.~\eqref{eq:SIC}, the ESIC criterion proposed in Ref.~\cite{Shang.etal2018} is recovered.
As demonstrated by various examples in Ref.~\cite{Shang.etal2018}, the ESIC criterion is more powerful as compared to CCNR.
So, here, we take a step further by asking the question whether the criterion as defined in Proposition~\ref{pro1} can be improved again if quantum $2$-designs besides SIC POVMs are utilized.
To distinguish the ESIC criterion, we dub the one using quantum $2$-designs as E$2$D; see below. 
\begin{proposition}[E$2$D]
  Let $\{\Pi^A\}$ and $\{\Pi^B\}$ be the normalized $2$-designs acting on the bipartite state $\rho_{AB}$.
  Then the linear correlation matrix is
\begin{equation}
  [\cC_\Pi]_{ij}=\bigl\langle \Pi_{i}^A\otimes\Pi_{j}^B\bigr\rangle=\tr\bigl(\rho_{AB}\Pi_{i}^A\otimes \Pi_{j}^B\bigr)\,.
\end{equation}
If $\rho_{AB}$ is separable, then
  \begin{equation}
    ||\cC_\Pi||_{\tr}\leqslant 1
  \end{equation}
  has to hold; Otherwise, it is entangled.
\end{proposition}
The proof is similar to that for Proposition~\ref{pro1}, which can be found in Refs.~\cite{Shang.etal2018,SarbickiGniewomir2020}.

\subsection{Nonlinear criteria}
The LUR criterion proposed in Ref.~\cite{Otfried2006} makes use of LOOs $\{G_k^A\}$ and $\{G_k^B\}$, such that for separable states,
\begin{equation}\label{eq:LUR}
  1-\sum_k\langle G_k^{A}\otimes G_k^{B}\rangle-\frac{1}{2}\sum_k\langle G_k^{A}\otimes \openone-\openone\otimes G_k^{B}\rangle^2\geqslant0
\end{equation}
has to hold.
One notices that LUR is strictly stronger than CCNR due to the extra nonlinear quadratic term in Eq.~\eqref{eq:LUR}.
In view of this fact as well as the fact that ESIC is stronger than CCNR, we try to reformulate the LUR criterion by replacing the LOOs with the normalized SIC POVMs.
See the following proposition which we dub as the LSIC criterion.
The detailed derivation is postponed in Appendix~\ref{App:proofLSIC}.
\begin{proposition}[LSIC]\label{LSIC}
Let $\{E_k^A\}$ and $\{E_k^B\}$ be the normalized SIC POVMs acting on two subsystems.
If a bipartite state $\rho_{AB}$ is separable, then
\begin{equation}\label{eq:LSIC}
  1+\sum_k\langle E_k^A\otimes E_k^B\rangle-\frac{1}{2}\sum_k\langle E_k^A\otimes\openone+\openone\otimes E_k^B\rangle^2\geqslant0
\end{equation}
has to hold; Otherwise, it is entangled.
\end{proposition}
Notice the sign differences in Eq.~\eqref{eq:LSIC} as compared to Eq.~\eqref{eq:LUR}.
Similarly, here, we are interested in the question whether the criterion as defined in Proposition~\ref{LSIC} can be further improved if quantum $2$-designs besides SIC POVMs are utilized.
To distinguish the LSIC criterion, we dub the one with quantum $2$-designs as L$2$D; see below. 
\begin{proposition}[L$2$D]
  Let $\{\Pi_k^A\}$ and $\{\Pi_k^B\}$ be the normalized $2$-designs acting on two subsystems.
  If a bipartite state $\rho_{AB}$ is separable, then
  \begin{equation}\label{eq:LSIC}
    1+\sum_k\langle \Pi_k^A\otimes \Pi_k^B\rangle-\frac{1}{2}\sum_k\langle \Pi_k^A\otimes\openone+\openone\otimes \Pi_k^B\rangle^2\geqslant0
  \end{equation}
  has to hold; Otherwise, it is entangled.
\end{proposition}
The proof is similar to that for Proposition~\ref{LSIC}, which can be found in Appendix.~\ref{App:proofLSIC}.

\section{Applications}\label{sec:Appl}
In this section, we test various entanglement criteria proposed above using simple $2\times2$, $3\times3$, and $2\times3$ entangled quantum states.

\subsection{$2\times2$ entangled states}
For the first application, we consider the noisy $2\times2$ quantum states with the form \cite{Otfried2006}
\begin{equation}\label{eq:rho2qb}
  \rho_{\rm{2qb}}(p)=p\ket{\psi}\bra{\psi}+(1-p)\rho_{s}\,,
\end{equation}
where the entangled state $\ket{\psi}$ can be set to be one of the Bell states,
\begin{align}
    \ket{\psi^{\pm}}&=\frac1{\sqrt{2}}\bigl(\ket{01}\pm\ket{10}\bigr)\,,\\
    \ket{\phi^{\pm}}&=\frac1{\sqrt{2}}\bigl(\ket{00}\pm\ket{11}\bigr)\,,
\end{align}
and the separable noise $\rho_{s}$ is given by
\begin{equation}
  \rho_{s}=\frac{2}{3}\ket{00}\bra{00}+\frac1{3}\ket{01}\bra{01}\,.
\end{equation}
Using the PPT criterion, these four families of states can be checked to be entangled for any ${p>0}$.

The number of elements of the quantum $2$-designs that we choose for testing are ${N=4,7,9}$ respectively.
Note that when ${N=4}$, it is simply the SIC POVM.
Table~\ref{tab:2qb} shows the threshold values of $p$ reported by various criteria.
The smaller the threshold is, the better the corresponding criterion is.
Several other features can be observed from the table.
First, ESIC is exactly equivalent to E$2$D, so is the pair of LSIC and L$2$D.
In other words, there is no difference of the detection power between the ESIC and E$2$D criteria, as well as the LSIC and L$2$D criteria.
This tells us that using quantum $2$-designs with more settings like ${N=7,9}$ is not helpful for improving the detection power as compared to SIC POVM with ${N=4}$.
With this simple example, we provide additional evidence for the potentially unique role played by SIC POVMs in quantum information processing \cite{appleby2017}.
Second, except for the case of $\ket{\psi^{+}}$, LUR performs better than ESIC. 
It is worth noting that although both criteria have been proposed previously, the comparison between them is missing until now.
This observation, to some extent, refutes our intuition that nonlinear criteria such as LUR are always better than linear ones like ESIC.
Next, both the EISC and LUR criteria are better than CCNR.
Finally, the LSIC and L$2$D criteria are completely ineffective for detecting certain entangled states.
This observation proves again that the advantage of nonlinear criteria over linear ones does not always exist.

\begin{table}[t]
\caption{\label{tab:2qb}Threshold values of $p$ for detecting the entangled states as in Eq.~\eqref{eq:rho2qb} using various criteria.\footnote{Note that the LOOs used in the LUR criterion are different for each case by invoking the Schmidt decomposition.} Since the states are entangled for any ${p>0}$ with PPT, the smaller the threshold is, the better the corresponding criterion is.}

\begin{tabular*}{\columnwidth}{@{\extracolsep{\fill}}cccccccccc}
\hline\hline
    && PPT & CCNR   & ESIC   & E$2$D  & LUR    & LSIC   & L$2$D  \\[2pt]
\hline
&$\ket{\psi^{-}}$ & $0$   & $0.2918$ & $0.2678$ & $0.2678$ & $0.2501$ & $0.2501$ & $0.2501$ &\\[2pt]
&$\ket{\psi^{+}}$& $0$   & $0.2918$ & $0.2678$ & $0.2678$ & $0.2779$ & $1$      & $1$   &\\[2pt]
&$\ket{\phi^{\pm}}$ & $0$   & $0.2164$ & $0.2053$ & $0.2053$ & $0.2028$ & $1$      & $1$  &\\[2pt]
\hline\hline
\end{tabular*}%
\end{table}
\begin{table}[h]
\caption{\label{tab:2qbrandom}For the $50\,000$ randomly generated $2\times2$ entangled states, the values in the table show the proportions that can be detected by various criteria.}
\begin{tabular*}{\columnwidth}{@{\extracolsep{\fill}}ccccccccc}
\hline\hline
&PPT & CCNR & ESIC & E$2$D & LUR & LSIC & L$2$D &\\[2pt]
\hline
&$100\%$ & $86.39\%$ & $88.52\%$ & $88.52\%$ & $87.48\%$ & $3.86\%$ & $3.86\%$ &\\[2pt]
\hline\hline
\end{tabular*}
\end{table}

To go a step further, we generate random two-qubit states according to the Haar measure and keep $50\,000$ entangled ones that are NPT, then check the proportions that can be detected by various criteria; see the results in Table~\ref{tab:2qbrandom}.
Apart from the same features that we can draw as those in Table~\ref{tab:2qb}, we find that ESIC and E$2$D are able to detect more states as compared to LUR and CCNR.
The LSIC and L$2$D criteria, however, are the weakest among all.
Moreover, all the states that can be detected by LSIC and L$2$D can also be detected by all the other criteria.
Here, we emphasize that SIC POVMs may play a fundamental role for detecting entanglement considering the criteria that we investigate in this work.
Experimentally, this special feature automatically provides the minimal number of settings that one should choose for the task of entanglement detection.

\subsection{$3\times3$ entangled states}
We move on the consider the ${3\times3}$ entangled quantum states.
In dimension ${d=3}$, there exist three different families of SIC POVMs, from which we arbitrarily choose one for testing.
For each SIC POVM, the number of elements is given by ${N=9}$.
For other quantum $2$-designs that we employ for the E$2$D and L$2$D criteria, we superimpose one set of SIC POVM over another (with proper rotations) to get a measurement with ${N=18}$ elements.
Note that the arbitrariness in choosing quantum $2$-designs is ensured by its rotational symmetry \cite{Ambainis.etal2007}.

\subsubsection{Bound entangled states}
We first consider the ${3\times3}$ bound entangled states \cite{Bennett1999} mixed with white noise
\begin{equation}\label{eq:BE}
  \rho(p)=p\rho_{\rm{BE}}+(1-p)\frac{\openone}{9}\,,
\end{equation}
where
\begin{equation}
  \rho_{\rm{BE}}=\frac1{4}\!\left(\openone-\sum_{i=0}^4\ket{\psi_i}\bra{\psi_i}\right)\!,
\end{equation}
with
\begin{align}
  &\ket{\psi_0}=\frac1{\sqrt{2}}\ket{0}(\ket{0}-\ket{1})\,,\quad \ket{\psi_1}=\frac1{\sqrt{2}}(\ket{0}-\ket{1})\ket{2}\,,\nonumber\\
  &\ket{\psi_2}=\frac1{\sqrt{2}}\ket{2}(\ket{1}-\ket{2})\,,\quad \ket{\psi_3}=\frac1{\sqrt{2}}(\ket{1}-\ket{2})\ket{0}\,,\nonumber\\
  &\ket{\psi_4}=\frac1{3}(\ket{0}+\ket{1}+\ket{2})(\ket{0}+\ket{1}+\ket{2})\,.
\end{align}

Table~\ref{tab:BE} shows the threshold values of $p$ reported by various criteria.
Similar to Table~\ref{tab:2qb}, the smaller the threshold is, the better the corresponding criterion is.
One finds that ESIC and E$2$D are equivalent, which are better than LUR, and all of them are better than CCNR.
For this particular entangled state, the LSIC and L$2$D criteria are completely ineffective.

\begin{figure}[t]
  \includegraphics[width=.96\columnwidth]{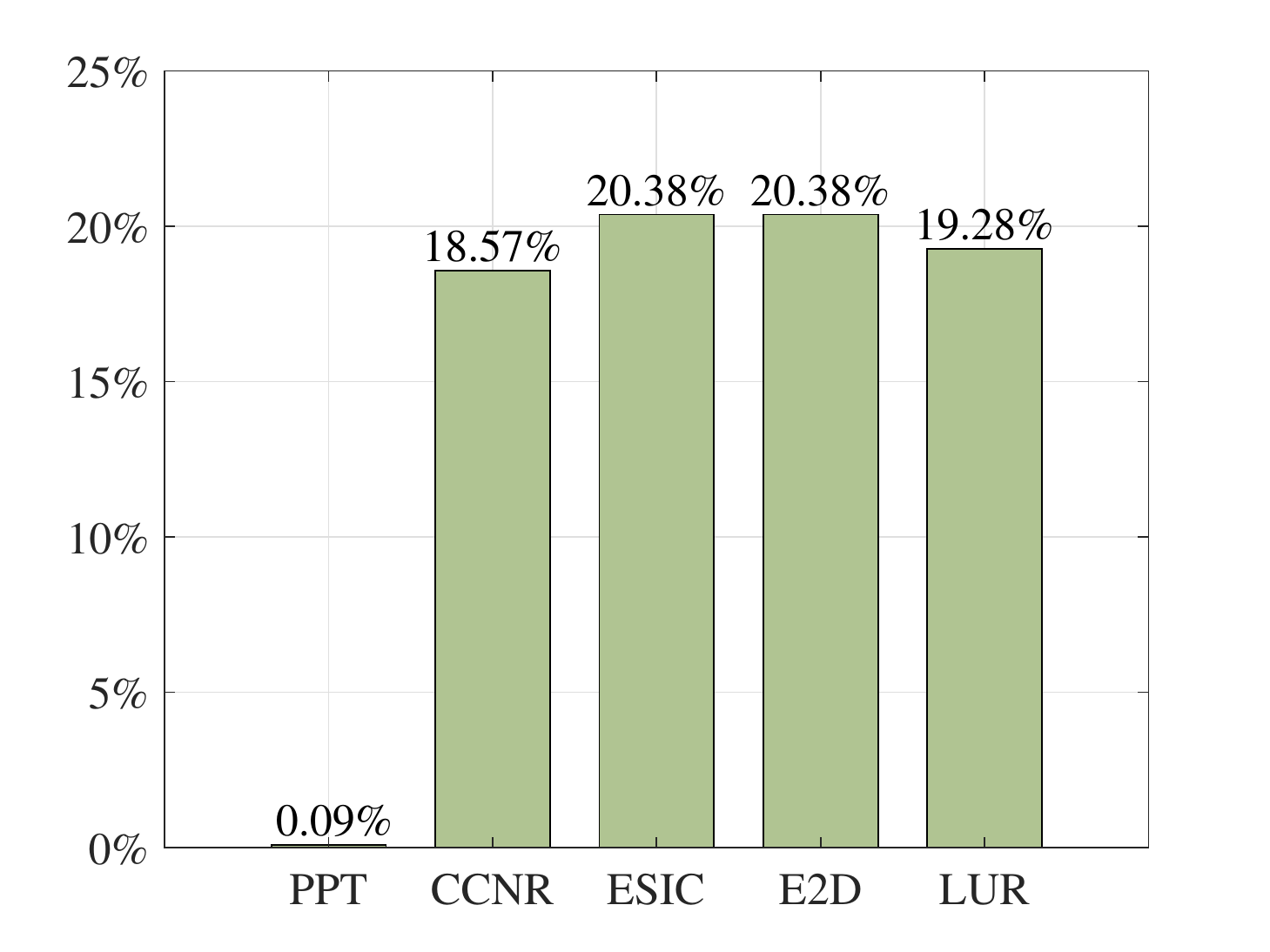}
  \caption{\label{fig:chessboard}Entanglement detection of the ${3\times3}$ chessboard states.
  For the $50\,000$ randomly generated states, the plot shows the fractions that are detected by various criteria.
  Within numerical fluctuations, the PPT criterion fails completely.
  The ESIC and E$2$D criteria can detect roughly $1\%$ more states than that of LUR, and roughly $2\%$ more than that of CCNR.
  }
\end{figure}
\begin{table}[h]
\caption{\label{tab:BE}  Threshold values of $p$ for detecting the $3\times3$ bound entangled states mixed with white noise as in Eq.~\eqref{eq:BE} using various criteria. The smaller the threshold is, the better the corresponding criterion is.}
\begin{tabular*}{\columnwidth}{@{\extracolsep{\fill}}ccccccccc}
\hline\hline
 &PPT & CCNR & ESIC & E$2$D & LUR & LSIC & L$2$D &\\[2pt]
\hline
 &$1$ & $0.8897$ & $0.8844$ & $0.8844$ & $0.8885$ & $1$ & $1$ &\\[2pt]
\hline\hline
\end{tabular*}
\end{table}
\subsubsection{Chessboard states}
Next we consider the ${3\times3}$ chessboard states defined as \cite{PhysRevA.61.030301}
\begin{equation}
  \rho_{\rm{chess}}=\mathcal{N}\sum_{j=1}^4|V_j\rangle\langle V_j|\,,
\end{equation}
where $\mathcal{N}$ is the normalization coefficient and the unnormalized vectors $\ket{V_j}$ are
\begin{align}
\ket{V_1} &=\ket{v_5,0,v_1v_3/v_6;0,v_6,0;0,0,0}\,,\nonumber\\
\ket{V_2} &=\ket{0,v_1,0;v_2,0,v_3;0,0,0}\,,\nonumber\\
\ket{V_3} &=\ket{v_6,0,0;0,-v_5,0;v_1v_4/v_5,0,0}\,,\nonumber\\
\ket{V_4} &=\ket{0,v_2,0;-v_1,0,0;0,v_4,0}\,.
\end{align}
We randomly generate $50\,000$ chessboard states with the six parameters $v_k$s taking values independently from a Gaussian distribution with standard deviation two and mean zero.
Figure~\ref{fig:chessboard} illustrates the fractions of states that are detected by various criteria.
Within numerical fluctuations, we find that the PPT criterion fails completely.
Again, the ESIC and E$2$D criteria are exactly equivalent.
They can detect roughly $1\%$ more states than that of LUR, and roughly $2\%$ more than that of CCNR.
Moreover, the LSIC and L$2$D criteria can hardly detect any chessboard states, thus are not shown in the figure.

\subsubsection{Horodecki states}
\begin{figure}[t]
  \includegraphics[width=.96\columnwidth]{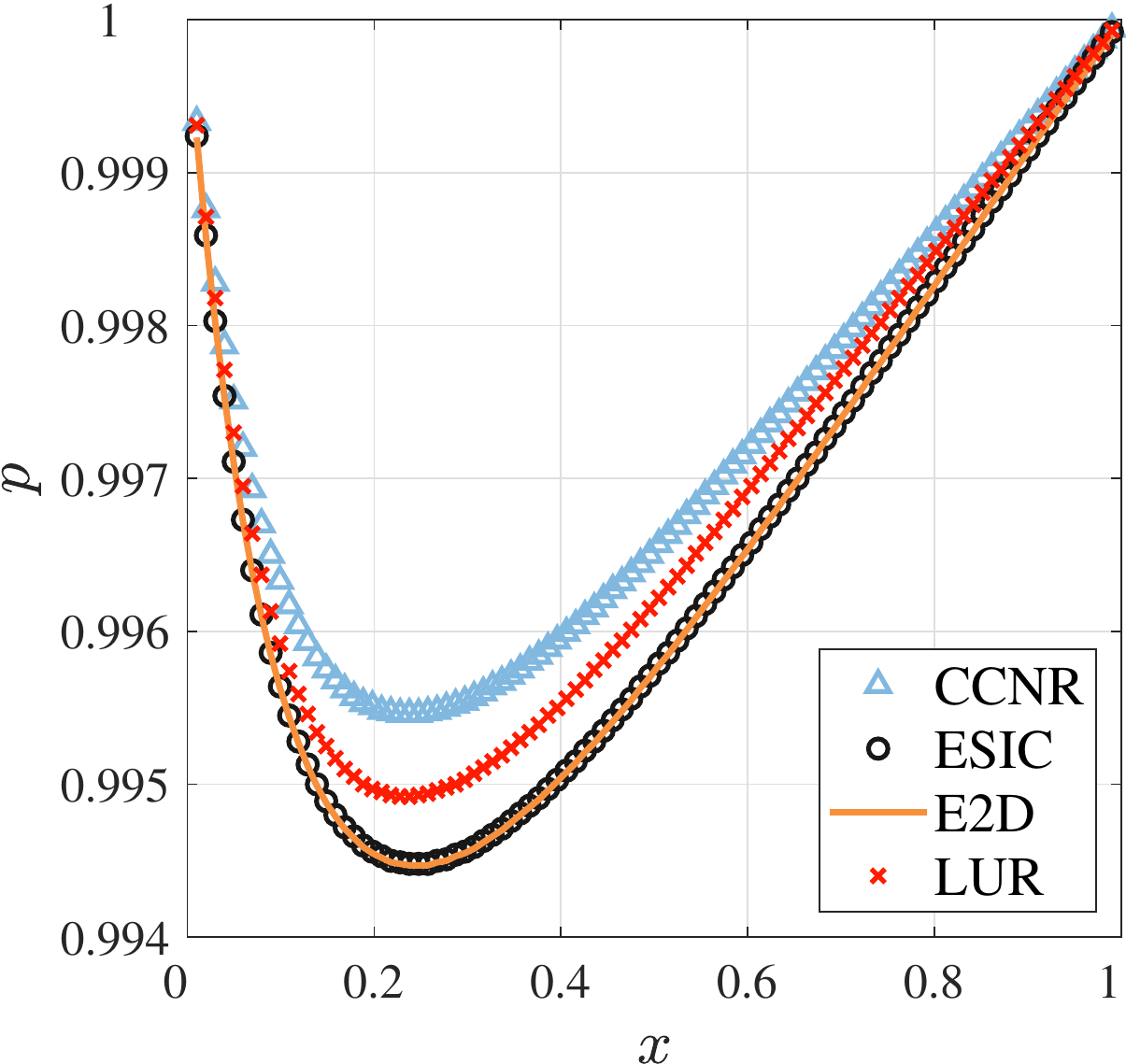}
  \caption{\label{fig:Horodecki}Entanglement detection of the ${3\times3}$ bound entangled Horodecki states mixed with white noise as in Eq.~\eqref{eq:PH}.
  States above the curves can be detected by the corresponding criterion.
  The ESIC and E$2$D criteria are exactly equivalent, and they are better than LUR.
  The CCNR criterion is the worst among all.
  }
\end{figure}

The ${3\times3}$ bound entangled states introduced by Horodecki are given by \cite{HORODECKI1997333}
\begin{equation}
\rho_{\rm{PH}}^x=
\frac{1}{8x+1}\!\left(\!\begin{array}{ccccccccc}
        x & 0 & 0 & 0 & x & 0 & 0 & 0 & x \\
        0 & x & 0 & 0 & 0 & 0 & 0 & 0 & 0 \\
        0 & 0 & x & 0 & 0 & 0 & 0 & 0 & 0 \\
        0 & 0 & 0 & x & 0 & 0 & 0 & 0 & 0 \\
        x & 0 & 0 & 0 & x & 0 & 0 & 0 & x \\
        0 & 0 & 0 & 0 & 0 & x & 0 & 0 & 0 \\
        0 & 0 & 0 & 0 & 0 & 0 & \frac{1+x}{2} & 0 & \frac{\sqrt{1-x^2}}{2} \\
        0 & 0 & 0 & 0 & 0 & 0 & 0 & x & 0 \\
        x & 0 & 0 & 0 & x & 0 & \frac{\sqrt{1-x^2}}{2} & 0 & \frac{1+x}{2} \\
\end{array}\!\right)\!,
\end{equation}
with the parameter ${0< x < 1}$.
Although these states cannot be detected by the PPT criterion and are not distillable, they are nevertheless all entangled.
Consider the mixture of $\rho_{\rm{PH}}^x$ with white noise
\begin{equation}\label{eq:PH}
\rho(x,p)=p\rho_{\rm{PH}}^x+(1-p)\frac{\openone}{9}\,,\quad 0\leqslant p\leqslant 1\,.
\end{equation}
In Fig.~\ref{fig:Horodecki}, we show the parameter ranges that are detected by various criteria.
All the states above the curves can be detected by the corresponding criterion.
One finds that the ESIC and E$2$D criteria are exactly equivalent, and both of them are better than LUR.
The CCNR criterion is the worst among all.
Moreover, the LSIC and L$2$D criteria can hardly detect any states, thus are not shown in the figure.

\subsection{$2\times3$ entangled states}
For the last application, we consider the $2\times3$ entangled states that are NPT.
In this case, the LUR, LSIC, and L$2$D criteria simply do not apply since all of them require the balanced dimension of the two subsystems.
We randomly generate $50\,000$ entangled states according to the Haar measure, and the values shown in Table.~\ref{tab:2by3} represent the proportions that can be detected by various criteria.
Once more, we confirm that the ESIC criterion is exactly equivalent to the E$2$D criterion, and they are all better than CCNR.

\begin{table}[h]
\setlength{\tabcolsep}{3mm}
\caption{\label{tab:2by3}For the $50\,000$ randomly generated $2\times3$ entangled states, the values in the table show the proportions that can be detected by various criteria.}
\begin{tabular*}{\columnwidth}{@{\extracolsep{\fill}}cccccc}
\hline\hline
&PPT     & CCNR    & ESIC    & E$2$D &\\[2pt]
\hline
&$100\%$ & $38.13\%$ & $41.62\%$ & $41.62\%$ &\\[2pt]
\hline\hline
\end{tabular*}
\end{table}
%

\section{Summary}\label{sec:Summary}
From an operational point of view, we constructed several new entanglement criteria in this work.
We first generalized the ESIC criterion \cite{Shang.etal2018} to the E$2$D criterion using quantum $2$-designs.
Then, based on the LUR criterion \cite{Hofmann2003,Otfried2006}, we proposed the LSIC criterion using SIC POVMs, and more generally the L$2$D criterion using quantum $2$-designs. 
Counter-intuitively, the E$2$D and L$2$D criteria with more settings are exactly equivalent to the corresponding ESIC and LSIC criteria respectively. 
In other words, there is no difference of the detection power between the ESIC and E$2$D criteria, as well as the LSIC and L$2$D criteria.
Fundamentally, this observation highlights the potentially unique role played by SIC POVMs in quantum information processing.
Experimentally, this provides the minimal number of settings that one should choose for entanglement detection.
Moreover, we find that nonlinear criteria such as LUR are not always better than linear ones like ESIC. 
This refutes the previous intuition that nonlinear criteria are always  better than linear ones \cite{Shang.etal2018}. 

As an outlook, it is interesting to re-investigate the corresponding criteria using quantum $t$-designs with ${t > 2}$ for multipartite entanglement detection,
for which we would also expect no improvement over that of the simplest setting of SIC POVMs.

\acknowledgments
We are grateful to Ali Asadian and Huangjun Zhu for helpful discussions.
This work was supported by the National Natural Science Foundation of China (Grants No.~12175014 and No.~11805010).

\appendix

\section{Proof of Proposition~\ref{LSIC}}\label{App:proofLSIC}
Here we explicitly derive the LSIC criterion as introduced in Proposition~\ref{LSIC}.
Consider the normalized SIC POVM in a finite dimension $d$ as in Eq.~\eqref{eq:SIC}, i.e.,
\begin{equation}
  E_k=\sqrt{\frac{d+1}{2d}}\ket{\psi_k}\bra{\psi_k}\,,
\end{equation}
and the square of it is given by
\begin{equation}
E_k^2=\frac{d+1}{2d}\ket{\psi_k}\bra{\psi_k}\,.
\end{equation}
Because of the completeness condition, we have
\begin{equation}
\sum_{k} E_k^2=\frac{d+1}{2}\openone\,.
\end{equation}
The corresponding probabilities satisfy the relation
\begin{equation}
  \sum_{k}e_k^2\leqslant 1\,,
\end{equation}
where
\begin{equation}
  e_k=\tr(\rho E_k)=\langle E_k \rangle
\end{equation}
is the probability for getting the $k$th outcome.

Variance of the measurement operator is given by
\begin{align}
\sum_{k}\varDelta ^2(E_k)
&=\sum_k\langle E_k^2\rangle-\sum_k\langle E_k\rangle^2\nonumber\\
&=\frac{d+1}{2}\langle\openone\rangle-\sum_k e_k^2\nonumber\\
&\geqslant\frac{d-1}{2}\,.
\end{align}
Next, we discuss the properties of local uncertainty relations \cite{Hofmann2003,Otfried2006}. 
In general, a pair of quantum systems $A$ and $B$ can be characterized by the operators $A_k$ and $B_k$ with the sum uncertainty relations giving by
\begin{equation}\label{LURpro}
\sum_{k}\varDelta ^2(A_k)\geqslant C_A, \quad \sum_{k}\varDelta ^2(B_k)\geqslant C_B\,.
\end{equation}
The positive values $C_A$ and $C_B$ can be computed by considering all the states. 
For separable states, the measurement outcomes are uncorrelated, so the variance of $A_k\otimes\openone+\openone\otimes B_k$ is equal to the sum of the local variances, i.e.,
\begin{equation}
\sum_{k}\varDelta ^2(A_k\otimes\openone+\openone\otimes B_k)\geqslant C_A+C_B\,.    
\end{equation}
Then, we have
\begin{equation}
\sum_{k}\varDelta ^2(E_k^A\otimes\openone+\openone\otimes E_k^B)\geqslant 2\times \frac{d-1}{2}=d-1\,.   
\end{equation}
Meanwhile,
\begin{align}
&\sum_{k}\varDelta ^2(E_k^A\otimes\openone+\openone\otimes E_k^B)\nonumber\\ 
=&\sum_k\langle E_k^2\otimes\openone+\openone\otimes E_k^2+2(E_k^A\otimes E_k^B)\rangle\nonumber\\ 
&\quad-\sum_k\langle E_k^A\otimes\openone+\openone\otimes E_k^B\rangle^2\nonumber\\ 
=&d+1+2\sum_k\langle E_k^A\otimes E_k^B\rangle-\sum_k\langle E_k^A\otimes\openone+\openone\otimes E_k^B\rangle^2\,.\nonumber\\
\end{align}
Combining the above two equations, one obtains
\begin{equation}
1+\sum_k\langle E_k^A\otimes E_k^B\rangle-\frac{1}{2}\sum_k\langle E_k^A\otimes\openone+\openone\otimes E_k^B\rangle^2\geqslant0\,,
\end{equation}
which recovers Eq.~\eqref{eq:LSIC} in the main text.

%

%

\end{document}